\documentclass[prb,showpacs,preprintnumbers,showkeys,amsmath,amssymb,preprint]{revtex4}
\usepackage{graphicx}
\usepackage{epsfig}
\usepackage{bm}

\begin{document}

\title[Dzyaloshinskii-Moriya interaction]{Detection and measurement of the
Dzyaloshinskii-Moriya interaction in double quantum dot systems}
\author{Sucismita Chutia}
\author{Mark Friesen}
\author{Robert Joynt}
\affiliation{Department of Physics, University of Wisconsin-Madison, Madison, WI
53706-1390}
\begin{abstract}
Spins in quantum dots can act as the qubit for quantum computation. In this
context we point out that spins on neighboring dots will experience an
anisotropic form of the exchange coupling, called the Dzyaloshinskii-Moriya
(DM) interaction, which mixes the spin singlet and triplet states. This will
have an important effect on both qubit interactions and spin-dependent
tunneling. We show that the interaction depends strongly on the direction of
the external field, which gives an unambiguous signature of this effect. We
further propose a new experiment using coupled quantum dots to detect and
characterize the DM interaction.
\end{abstract}
\volumeyear{year}
\volumenumber{number}
\issuenumber{number}
\eid{identifier}
\date{04/28/2006}
\received[Received text]{date}

\revised[Revised text]{date}

\accepted[Accepted text]{date}

\published[Published text]{date}

\maketitle

Solid state devices show great promise for scalable quantum information
processing. Several well known proposals for quantum computing have been
presented, including semiconducting quantum dots \cite{Loss98} and
superconducting Josephson junctions.\cite{Makhlin01}  Quantum dots currently
enable the confinement and control of electrons on the scale of tens of
nanometers, even down to the limit of one electron.\cite{Ciorga00,Elzerman03}
Detection techniques allow the measurement of a single electron spin.\cite{Hanson05}

The solid state matrix provides both opportunities and challenges for
quantum control and decoherence, due to the complex environment of the
qubits. In this paper, we focus on a prominent issue for many solid state
qubit implementations: the spin orbit interaction, which couples spin and
charge fluctuations. As typical for qubit interactions, the spin orbit
coupling can be both useful \cite{Wu02,Friesen04} and detrimental.\cite{phonons,tahan02}
Here, we consider how the spin orbit coupling affects the time
evolution of two-qubit interactions in spin-based quantum dot qubits. We
find that failure to account for spin orbit coupling can lead to serious
control errors in the quantum computation.

The main interaction between spin qubits is the exchange coupling, which can
be controlled with electronic gates, by raising or lowering the
electrostatic tunnel barrier between neighboring quantum dots \cite{Loss98} or by varying the relative depth of the wells constituting the double dot.\cite{Petta05} 
Ignoring the crystal matrix, the exchange coupling would be of the
Heisenberg type, with global $SU(2)$ spin symmetry: 
$H_{\text{Heis}}\ =J\mathbf{S}_{1}\cdot \mathbf{S}_{2}$. Here, $\mathbf{S}_{i}$ 
are spin operators and $J$ is
the tunable exchange coupling constant. The presence of spin-orbit interaction
introduces anisotropy into the exchange coupling, with an antisymmetric
component known as the Dzyaloshinskii-Moriya (DM) interaction.\cite{dzy,moriya}
Kavokin has shown that the DM exchange term also occurs in quantum dots, as a consequence of 
tunneling coupling.\cite{Kavokin01} Devitt \textit{et al.} have proposed methods to determine its magnitude.\cite{Devitt}  Several other authors have also studied the importance
of the DM interaction for quantum dot quantum 
computing.\cite{Wu02,Bonesteel01,Burkard02,Stepanenko03,Kavokin04}

For a single pair of dots, we can write the DM interaction as $H_{%
\text{DM}}\ =\alpha J\,\hat{\mathbf{r}}\cdot \left( \mathbf{S}_{1}\times \mathbf{S}%
_{2}\right) $, where $\hat{\mathbf{r}}$ is the unit vector joining the two spins.
Thus, the presence of a DM interaction reduces the spin symmetry to a global 
$U(1)$ cylindrical symmetry, where only rotations about $\hat{\mathbf{r}}$ remain
as symmetry operations. Kavokin has considered the magnitude of the
coefficient $\alpha ,$ computing the exchange integral for the two electrons
taking into account the admixture of spin projections caused by the
spin-orbit interaction.\cite{Kavokin01} 
This calculation applies to the case when the energy levels of the individual dots 
are approximately equal ($|\epsilon_1-\epsilon_2| \lesssim J$) and the Heitler-London method is valid.
Here, we consider only this particular case.  In GaAs, the
predominant spin-orbit coupling is of the Dresselhaus type. For quantum dots in a 100 \AA~ GaAs quantum well, Kavokin finds $\alpha\approx 0.1$, not a particularly small value. 
In such a case, we expect the DM contribution to the exchange coupling will be
readily apparent. For silicon dots, the Dresselhaus interaction is not
present, and the predominant spin-orbit coupling arises from the Rashba
interaction.

Petta \textit{et al.} have recently performed a set of experiments with
coupled spins in a double quantum dot system in GaAs that demonstrate control of the
exchange coupling.\cite{Petta05} In these experiments, a qubit was defined
by $|S\rangle $ and $|T_{0}\rangle $: the singlet, and one component of the
triplet states of the two-spin system respectively. Neglecting additional
couplings, we would expect the Heisenberg term term to split $|S\rangle $
and $|T_{0}\rangle $, thus enabling exchange-based qubit rotations.\cite{Petta05} 
However, inhomogeneous nuclear fields and the DM interaction also
mix in $|T_{+}\rangle $ and $|T_{-}\rangle $, the two other components of
the triplet. \ The resulting loss of wavefunction probability from the qubit
subspace constitutes leakage and it can be interpreted as dephasing or decoherence. However, 
the dynamics are actually coherent. It may therefore be possible to utilize the
DM dynamics in a beneficial way,\cite{Wu02,stepanenko04} or to undo them using 
time-symmetric pulse shapes \cite{Bonesteel01,Burkard02,Stepanenko03} or spin echo
techniques. Here, we investigate in detail the DM dynamics of a double
quantum dot system, specifically considering the experiments of Petta 
\textit{et al.} We explore how the DM interaction modifies the usual
interpretation of such experiments, and we propose further experiments to
detect the presence of the DM interaction and to measure its magnitude.

The Hamiltonian for our double-dot system is 
\begin{equation}
H=J~\mathbf{S}_{1}\cdot \mathbf{S}_{2}+\alpha J~\hat{\mathbf{r}}\cdot \left( \mathbf{S}%
_{1}\times \mathbf{S}_{2}\right) -g^{\ast }\mu _{B}\left( \mathbf{B}_{1}\cdot 
\mathbf{S}_{1}+\mathbf{B}_{2}\cdot \mathbf{S}_{2}\right) ,  \label{eq:H}
\end{equation}%
where $\mathbf{B}_{1}=\mathbf{B}_{\text{ext}}+\mathbf{B}_{n,1}$ and 
$\mathbf{B}_{2}=\mathbf{B}_{\text{ext}}+\mathbf{B}_{n,2}$. Here, $\mathbf{B}_{n,i}$ 
is the semiclassical field that is used to approximate the effective nuclear field for coupling of the electron spin to local nuclei in dot $i$,\cite{coish} given by
\begin{equation}
\mathbf{B}_{n,i}=\frac{A v_o}{-g^*\mu_B}\sum_k |\psi_0^i(\mathbf{r}_k)|^2 \mathbf{I}^k ,
\label{eq:Bn}
\end{equation}%
where $\mathbf{I}^k$ is the nuclear spin operator for a nucleus of total spin $I$ at the lattice site $k$, $v_0$ is the volume of a unit cell containing one nuclear spin, $A$ is the hyperfine coupling strength and $\psi_0^i (\mathbf{r}_k)$ 
is the single particle envelope function for the orbital state $i$ evaluated at site $k$. We assume that $\mathbf{B}_{n,i}$ has a Gaussian 
distribution with mean zero, and a typical variance of $\sigma=2.3$~mT.
In the calculations reported below, our results are
averaged over the distribution of the nuclear fields,\cite{coish} as consistent
with experimental the procedure.\cite{Petta05}  For example,
\begin{eqnarray}
\bar{F}=(2\pi \sigma^2)^{-3/2}\int_0^{2\pi}\! d\phi \int_0^{\pi} \! d\theta
\sin\theta\times\nonumber\\ \int_0^{\infty}\! dB_{n}\, B_{n}^2 e^{-B_{n}^2/2\sigma^2} 
F(\mathbf{B}_n) . \label{eq:Intg}
\end{eqnarray}
High dimensional integrals are evaluated numerically using a simple Monte Carlo 
integration code.

In Fig.~1, we show the appropriately averaged eigenvalues of $H$,
as a function of the applied field $B_{\text{ext}}$. 
(Note that we use $\alpha =0.5$ in this figure. \ This large value
of $\alpha $ is chosen only for purposes of illustration. Elsewhere in the
paper we use the more physical value $\alpha =0.1$.) 
We observe mixing of the unperturbed singlet and triplet states at
special fields. Near $B_{\text{ext}}=0$, there is mixing of the triplet
states, primarily due to inhomogeneous nuclear fields \cite{Petta05}. At nonzero
fields, there is an additional mixing of the singlet and triplet states,
which arises from both inhomogeneous nuclear fields and the DM interaction.
The mixing occurs near the resonance condition $g^*\mu_B B_{\text{ext}}\approx \pm J$,
corresponding to $B_{\text{ext}}\approx \pm 0.04$~T in the figure. From the
point of view of experimental detection, a crucial point is that the
mixing effect is anisotropic. This is seen clearly in Fig.~1(b) where we plot the overlaps between
the eigenstates of Eq.~(\ref{eq:H}) and the pure spin singlet.

\begin{figure}[t]
\begin{center}
\includegraphics[width=3in]{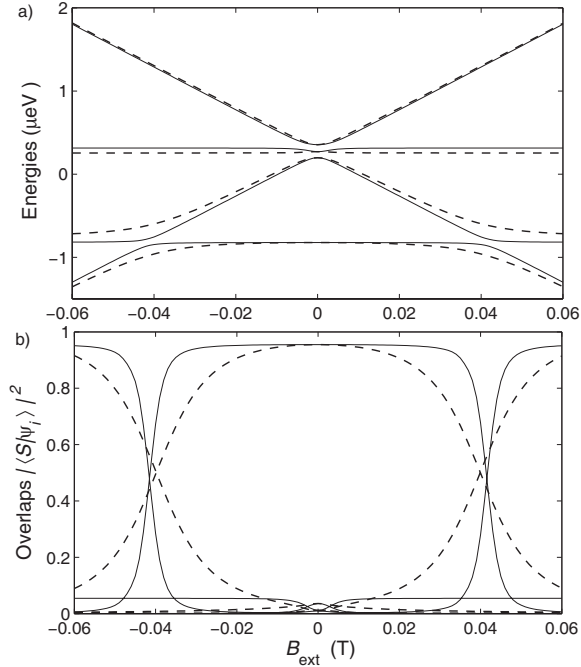}
\end{center}
\caption{ (a) Energy eigenvalues of the spin Hamiltonian, Eq.~(\protect\ref{eq:H}), 
 as a function of $B_{\text{ext}}$, for the parameters $J=1\,\protect\mu$eV, $\alpha=0.5$
and $\sigma=2.3$~mT .
Here the solid line corresponds to $\mathbf{B}_{\text{ext}} \| \hat{\mathbf{r}}$ 
and the dashed line corresponds to ${%
\ \mathbf{B}}_{\text{ext}} \perp \hat{\mathbf{r}}$. (b) The overlap $|\langle S | 
\protect\psi_i\rangle|^2$ of the energy eigenstates with the spin singlet as a
function of field. Solid and dashed lines have the same meaning as in (a).}
\end{figure}

We now compute the time evolution of the two coupled spins for several
experimental situations of interest. The Hamiltonian in Eq.~(\ref{eq:H}) has
four eigenstates $|\psi_{1..4}\rangle$, with the corresponding eigenvalues $%
E_{1..4}$. For an arbitrary initial state given by $|\Psi (0)\rangle =\sum
a_i |\psi_i \rangle$, we can compute the probability $P_s(t)=|\langle S |
\Psi(t) \rangle |^2$ that this state will evolve to a spin singlet after
time $t$. We consider the initial state $|\Psi (0)\rangle = |S\rangle$. The
probability that the spin system will remain in its singlet state is then given by 
\begin{equation}
P_s(t)=\sum_{i=1}^{4}|a_i|^4+ 2\sum_{i<j}|a_i|^2 |a_j|^2 \cos [ (E_i-E_j) t/
\hbar ] .  \label{eq:Ps}
\end{equation}

Leakage can occur due to both the DM interaction and the inhomogeneous
nuclear fields. When $J$ is exponentially suppressed, leakage is due
entirely to the nuclear fields. For non-vanishing $J$, the initial singlet
state would remain stationary if not for the nuclear and DM mechanisms. Both
mechanisms then play a role in leakage. In Fig.~2, we plot $P_{s}(t)$
obtained after allowing the system to evolve over a \textquotedblleft
waiting time" $t=\tau_s$. At the resonance condition $J=g^{\ast }\mu
_{B}B_{\text{ext}}$, $P_{s}$ is strongly suppressed compared to smaller and
larger fields. A similar suppression of $P_{s}$ is expected in the absence
of spin-orbit coupling. However, the DM relaxation mechanism exhibits a
strong dependence on the orientation of $\mathbf{B}_{\text{ext}}$ with respect
to $\hat{\mathbf{r}}$, which cannot be explained by nuclear fields. This
dependence on field orientation provides an important signature of the DM
interaction. 

\begin{figure}[t]
\begin{center}
\includegraphics[width=3 in]{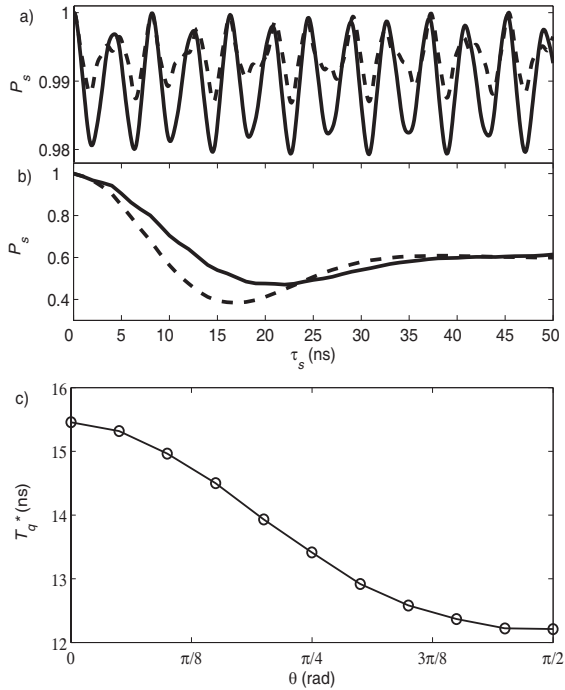}
\end{center}
\caption{ (a,b) The probability $P_s$ of an initial spin singlet to remain in
the singlet state as a function of the ``waiting time" $\protect\tau_s$ for
$J=1$ $\protect\mu$eV, $\alpha=0.1$ and $\sigma=2.3$~mT. Solid lines correspond to the field orientation $\mathbf{B}_{\text{ext}}\| \hat{\mathbf{r}}$  and
dashed lines correspond to $\mathbf{B}_{\text{ext}}\perp \hat{\mathbf{r}}$. 
(a) $B_{\text{ext}}=100$~mT, 
(b) $B_{\text{ext}}=40$~mT (resonance condition),
(c) Dephasing time $T_q^*$, obtained from a fit to a parabolic equation as a
function of the orientation angle $\protect\theta$ between $\mathbf{B}_{\text{ext%
}}$ and $\hat{\mathbf{r}}$. The fitting data correspond to the curves $B_{\text{ext}}%
=40$~mT in (b).}
\end{figure}
In the experiments of Petta \textit{et al.}, an initial singlet state is
prepared with both electrons in a single quantum dot. The electrons are
subsequently separated into two dots while retaining their singlet
correlations. A waiting time ensues, consistent with the analysis presented
above, after which the singlet probability $P_{s}$ is measured. The data can
be fit using a semiclassical model,\cite{Schulten78} obtaining a dephasing
time of about $T_{2}^{\ast }\approx 10$~ns and effective static nuclear field 
$B_{\text{nuc}}=2.3$~mT. This $T_{2}^{\ast }$ is an ensemble averaged time for 
relaxation to the asymptotic value. More
relevant for quantum information is the short-time behavior, characterized by the 
quantity $T_q^*$, defined by $P_{s}(\tau )\cong 1-\left( \tau/T_{q}^{\ast }\right) ^{2}$. In Fig. 2(c),  $T_{q}^{\ast }$ is plotted as a function of $\theta $, the angle between $\mathbf{B}_{\text{ext}}$
and $\hat{\mathbf{r}}$. For $B_{\text{ext}}=40 $~mT, the results show a
significant dependence on $\theta $. Note that the $t^2$ dependence of $P_s$ 
(as opposed to an exponential decay) is due to an absence of dissipation in our model.

The significance of the DM interaction becomes most apparent during exchange
gate operations, when the Heisenberg and DM couplings, $J$ and $\alpha J$
respectively, are non-vanishing. We consider the \textquotedblleft Rabi
oscillation" experiment of Petta \textit{et al.}, in which the spins are
initially prepared in the state $|n\rangle $, corresponding to
the ground state determined by the nuclear fields when $J=0$. The initial
state is not an eigenstate of $H_{\text{Heis}}$, so when the Heisenberg
interaction is initiated, coherent oscillations will occur between the
singlet and triplet manifolds. Thus, after an exchange period of $\tau
_{E}=2\pi \hbar /J$, the spins will return to their initial state. Both
inhomogeneous nuclear fields and the DM interaction affect this picture by
mixing in the different triplet states inhomogeneously, causing $%
P_{n}(t)=|\langle n|U(t)|n\rangle |^{2}$ to decay. Here, $P_{n}$ is the
probability to return to the initial state $|n\rangle $, \cite{note} and $%
U(t)$ is the unitary evolution operator for the spin Hamiltonian. If we
define the $a_{i}$ coefficients of the initial state as $|n\rangle =\sum
a_{i}|\psi _{i}\rangle $, then $P_{n}(t)$ is given by Eq.~(\ref{eq:Ps}).

We have computed $P_{n}(t)$ for experimental parameters consistent with 
Ref.~[\onlinecite{Petta05}]. The results are shown in Fig.~3. Here, the external field is
much larger than the nuclear field, so the initial state of the evolution is
nearly spin polarized. The exchange coupling is then switched suddenly to a
value slightly off from the resonant condition $J=g^{\ast }\mu _{B}B_{\text{%
ext}}$ for a period $\tau _{E}$. We plot two cases: with and without the DM
interaction ($\alpha =0.1,0$, respectively). In both cases, the initial
state $|\Psi (0)\rangle \approx |T_{+}\rangle $ is very similar to the
ground eigenstate. (Recall that the nuclear fields and the DM interactions
cause a hybridization of the $|S\rangle $ and $|T_{+}\rangle $ states near
their level crossing. But away from the crossing, the eigenstates retain
their $|S\rangle $ and $|T_{+}\rangle $ character.) Therefore, in the
long-time limit $\tau _{E}\gg \tau _{\text{nuc}}$, $P_{n}$ does not deviate
greatly from 1. Here, $\tau _{\text{nuc}}\approx \hbar /g^{\ast }\mu _{B}B_{%
\text{nuc}}$ is the nuclear mixing time. We note that the solution including
the DM interaction is clearly distinguishable from the $\alpha =0$ case.
This is because the DM coupling enhances the hybridization of $|S\rangle $
and $|T_{+}\rangle $, and thus the difference between the initial and final
states.

Another obvious feature in Fig.~3 is the initial rapid oscillations of $%
P_{n} $. Since the initial spin state is not an eigenstate of the exchange
Hamiltonian, it can undergo coherent oscillations prior to nuclear mixing.
In the figure, the predominant oscillations occur between the $S$-like and $%
T_{+}$-like states, with an approximate energy splitting of $g^{\ast }\mu
_{B}B_{\text{ext}}-J$ and a corresponding oscillation period of $2\pi \hbar
/(g^{\ast }\mu _{B}B_{\text{ext}}-J)$. \ Note that without any true
damping mechanisms the curves are subject to Poincar\'{e} recurrence and
will return to 1.

\begin{figure}[t]
\begin{center}
\includegraphics[width=3 in]{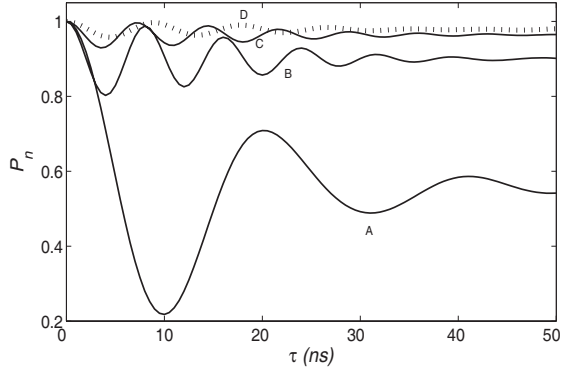}
\end{center}
\caption{ Coherent oscillations of $P_n$, corresponding to the ``Rabi
oscillations" of Ref.~[\protect\onlinecite{Petta05}] at $B = 100$~mT, 
and $\sigma=2.3$~mT. Solid curves
include the effects of the DM interaction. The dotted curve corresponds to
DM interactions turned off ($\protect\alpha=0$). (A) $J=2.5$~$\mu$eV, $\alpha=0.1$, (B)
$J=3$~$\mu$eV, $\alpha=0.1$, (C) $J=2$~$\mu$eV, $\alpha=0.1$, (D) $J=3$~$\mu$eV, $\alpha=0$. }
\end{figure}

We now propose an experiment to unambiguously detect the presence of the DM
interaction. In Fig.~3, the suppression of $P_{n}$ was strongly enhanced by
DM interactions near resonance (2.5 $\mu$eV). So we perform the previous experiment in a large external field where  we can tune the exchange coupling to
its resonant condition $J=g^{\ast }\mu _{B}B_{\text{ext}}$ during the
exchange evolution. Under these conditions, the hybridization of 
$|S\rangle $ and $|T_{+}\rangle $ is maximized, so that the initial and
final states will be quite different. Consequently, after nuclear mixing, $%
P_{n}$ approaches 0.5. The reason for choosing $B_{\text{ext}}$ (and thus $J$%
) to be large is that this allows many coherent oscillations to occur before
nuclear mixing.

\begin{figure}[t]
\begin{center}
\includegraphics[width=3 in]{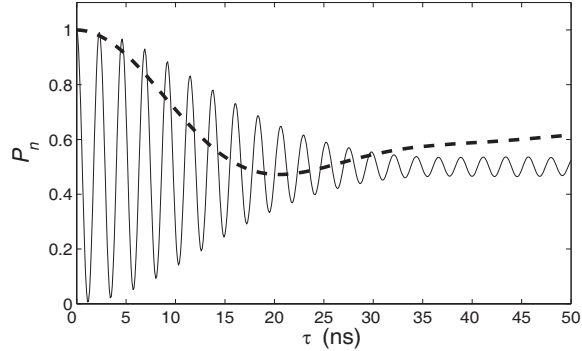}
\end{center}
\caption{ Proposed experiment to observe DM interactions. 
The system is prepared in the initial state $|T^+\rangle$ in a large field, 
$B_\text{ext}=1$~T. A strong exchange coupling ($J =
25.5$~$\protect\mu$eV) is initiated at the resonance condition $J=g^*\protect%
\mu_B B_{\text{ext}}$ with $\sigma=2.3$~mT, producing fast oscillations dampened by
nuclear mixing (solid curve). In the absence of the DM interaction, no fast
oscillations are observed (dashed curve).}
\end{figure}

Some typical results are shown in Fig.~4, with and without the DM
interaction. Because a large value of $J$ has been used, the hybridization
of $|S\rangle$ and $|T_+\rangle$ is completely dominated by the DM
interaction for the case $\alpha=0.1$. Rapid oscillations occur between
these two states, with an energy splitting given by $2|\langle T_+|H_{\text{DM}}|S\rangle| =\alpha J/\sqrt{2}$, and an oscillation period of $\pi \sqrt{8}%
\hbar/\alpha J$. To see the fast oscillations, the exchange coupling should
be turned on quickly compared to the oscillation period, so that the initial
state cannot evolve adiabatically to the ground state. Similar to Fig.~3,
the oscillation envelope is eventually suppressed by nuclear mixing.
However in the large $B_{\text{ext}}$ limit, the fast oscillation period is determined only by DM interactions, not hyperfine effects.  This can be confirmed by plotting
$[\text{(fast oscillation period)} \times B_{\text{ext}}]$ vs. $B_{\text{ext}}$, 
which should remain a
constant.  The hyperfine effects can also be eliminated by polarizing 
the nuclear spins or by employing a standard Hahn spin echo sequence.

Two-qubit operations require a very accurate knowledge of the spin-spin
interaction, and the DM interaction is expected to be about a 10\% effect in GaAs.
It is therefore very important to develop methods to measure it in double
quantum dot systems. Because the interaction breaks spin rotation
invariance it can be detected: its effects depend strongly on the direction
of the applied field in ways that we have described. By carefully choosing
external parameters, it is also possible to determine the magnitude of the
DM coupling by measuring the oscillation period for evolution between the singlet
and triplet states.

We gratefully acknowledge conversations with C. M. Marcus and W. O. Puttika. This work was supported by NSA and ARDA under ARO contract number W911NF-04-1-0389 and by the National Science Foundation through the ITR (DMR-0325634) and EMT (CCF-0523675) programs.

\end{document}